# OpenWSI: a low-cost, high-throughput whole slide imaging system via single-frame autofocusing and open-source hardware

Chengfei Guo,[1,4,6] Zichao Bian,[1,6] Shaowei Jiang,[1,6,*] Michael Murphy,[3] Jiakai Zhu,[1] Ruihai Wang,[1] Pengming Song,[2] Xiaopeng Shao,[1,4] Yongbing Zhang,[5] and Guoan Zheng,[1,2]

[1]Department of Biomedical Engineering, University of Connecticut, Storrs, CT, 06269, USA
[2]Department of Electrical and Computer Engineering, University of Connecticut, Storrs, CT, 06269, USA
[3]Department of Dermatology, University of Connecticut Health Center, Farmington, CT 06030, USA
[4]School of Physics and Optoelectronic Engineering, Xidian University, Shaanxi, 710071, China
[5]Graduate School at Shenzhen, Tsinghua University, Shenzhen, 518055, China
[6]These authors contributed equally to this work
*Corresponding author: shaowei.jiang@uconn.edu

**Recent advancements in whole slide imaging (WSI) have moved pathology closer to digital practice. Existing systems require precise mechanical control and the cost is prohibitive for most individual pathologists. Here we report a low-cost and high-throughput WSI system termed OpenWSI. The reported system is built using off-the-shelf components including a programmable LED array, a photographic lens, and a low-cost computer numerical control (CNC) router. Different from conventional WSI platforms, our system performs real-time single-frame autofocusing using color-multiplexed illumination. For axial positioning control, we perform coarse adjustment using the CNC router and precise adjustment using the ultrasonic motor ring in the photographic lens. By using a 20X objective lens, we show that the OpenWSI system has a resolution of ~0.7 μm. It can acquire whole slide images of a 225-mm² region in ~2 mins, with throughput comparable to existing high-end platforms. The reported system offers a turnkey solution to transform the high-end WSI platforms into one that can be made broadly available and utilizable without loss of capacity.**

Digital pathology via whole slide imaging (WSI) promises better and faster diagnosis and prognosis of cancers and other diseases [1]. A major milestone was accomplished in 2017 when the Philips' WSI system was approved for the primary diagnostic use in the US [2]. In a conventional WSI system, the tissue slide is mechanically scanned to different x-y positions and the digital images are acquired using a high numerical aperture (NA) objective lens. The small depth of field of the objective, however, poses a challenge for proper focusing during the scanning process [3, 4]. Many existing systems create a focus map prior to the scanning process. For each point on the map, the system needs to scan the sample to different axial positions and acquire a z-stack. The best focus position can then be inferred based on the image with the highest Brenner gradient or other figure of merits [5-7]. In this z-stack approach, surveying the focus points for every tile would require a prohibitive amount of time. Most systems select a subset of tiles for focus point surveying to save time. Different from the z-stack approach, we have recently demonstrated a focus map surveying method based on single-frame autofocusing [8, 9]. In this approach, we illuminate the sample from two different incident angles. If the object is placed at an out-of-focus position, the captured image would contain two copies of the object separated by a certain distance. The defocus distance can then be recovered based on the two-copy separation.

In this letter, we report the development of a low-cost, high-throughput DIY WSI system termed OpenWSI. The reported system is built using off-the-shelf components including a programmable LED array, a photographic lens, and a low-cost computer numerical control (CNC) router. Different from conventional platforms, our system does not perform focus map surveying. Instead, it performs real-time autofocusing in between two brightfield image acquisitions. Since the focus map is not needed in the scanning process, mechanical repeatability is not required in our design, enabling us to build a 3-axis scanning platform using a low-cost CNC router. Axial positioning control is critical for microscopy imaging. In the reported platform, we perform coarse axial adjustment using the CNC router and precise adjustment using the ultrasonic motor ring in the photographic lens. We also provide the implementation protocol on controlling the ultrasonic motor ring within a photographic lens. To the best of our knowledge, it is the first demonstration on how to employ a photographic lens for precise axial positioning. It will provide a simple yet powerful tool for 3D microscopy and time-lapse focus tracking. The cost of the reported system is affordable to individual researchers. This contribution is significant for biologists and microscopists. Experiments that were typically carried out manually in single-cell level and that addressed a limited field of view at a time can now be done for the entire sample in an automated manner. The result from the reported system is a comprehensive digital rending of the entire sample, on the order of centimeter in size, and visible at sub-micron resolution. Image analysis techniques, routines, and tools can then be for quantitative post-acquisition data analysis, which extracts important experimental information in a statistical manner.

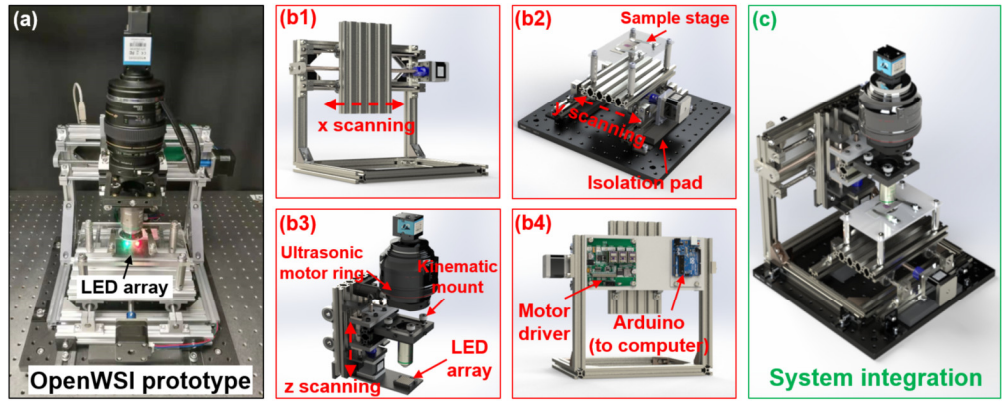

Fig. 1. (Visualization 1) OpenWSI: a low-cost, high-throughput whole slide imaging system with open-source hardware. (a) The OpenWSI prototype with a programmable LED array, a low-cost CNC router, and a Canon photographic lens. We perform real-time autofocusing using red / green LED illumination. Mechanical repeatability is not required in our design, allowing us to build the scanning platform under a $200 budget. The stages for mechanical scanning in the x- (b1), and y-directions (b2). (b3) Axial positioning using the CNC router and the ultrasonic motor ring. (b4) The CNC controller and the Arduino board for communication with the computer. (c) System integration.

Figure 1(a) shows the design of the OpenWSI system, where a Nikon 20X, 0.75 NA objective lens is used in this prototype setup. Instead of using a conventional microscope tube lens, we employ a Canon 100-mm photographic lens in our platform. This photographic lens allows us to perform precise axial positioning control and the cost is also lower compared to the conventional microscope tube lens. We use an 8 by 8 programmable LED array for sample illumination and a 20-megapixel camera for image acquisition (Sony IMX 183 sensor). The LED array allows us to switch between the regular brightfield imaging mode and the color-multiplexed autofocusing mode. Other more advanced microscopy techniques such as Fourier ptychographic microscopy, phase-contrast, and 3D tomographic imaging [10-13] can also be implemented using this design. The measured resolution of our prototype system is ~0.7 μm using a USAF target (resolving group 10, element 4 with a 0.35 μm half-pitch line width). Figure 1(b) shows the components in the OpenWSI design and Fig. 1(c) shows the system integration (Visualization 1). In our design, we modify a low-cost CNC router (Mysweety CNC router, Amazon) for 3D sample positioning. Figure 1(b1)-(b2) show the x- and y-scanning stages. In Fig. 1(b3), a kinematic mount (Thorlabs KC1-T) is used to hold the objective lens and enables precise angular alignment with respect to the sample stage. In Fig. 1(b4), the motor driver of CNC router is connected to an Arduino board that communicates with the computer via series commands.

One innovation of the reported system is to perform remote focus control at the image plane. Figure 2(a) shows the captured images by tuning the ultrasonic motor ring to different positions (Visualization 2). Figure 2(b) shows the measured calibration curve between the ring positions and the defocus positions of the sample. In this experiment, we use a precise mechanical stage (ASI LS-50) to mount the objective lens. For different ring motor positions, we use the precise mechanical stage to move the objective lens back to the in-focus position. Based on this calibration curve, we can see that one step of the lens ring rotation corresponds to an 8-μm axial shift at the image plane and an 80-nm shift at the object plane.

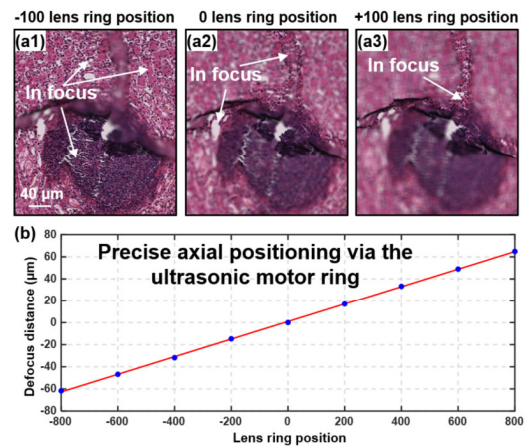

Fig. 2. (Visualization 2) Precise image-plane focus control via the ultrasonic motor ring within the photographic lens. (a) The captured images at different ring positions. (b) The measured calibration curve between the ring positions and the defocus positions.

There are two advantages of using this strategy for microscopy imaging. First, for many biomedical experiments, axially moving the stage or the objective lens may perturb the sample. The reported scheme is able to avoid potential mechanical perturbation during the experiment. Second, due to the magnification of the optical system, the precision needed for image-plane focus control is 100 times lower than that for object-plane focus control. It, thus, allows us to employ a cost-effective photographic lens for 3D imaging and precise focus tracking.

In the OpenWSI system, we perform single-frame autofocusing using a programmable LED array for sample illumination. We note that no condenser lens is needed in between the LED array and the sample. For regular brightfield imaging, the illumination NA of the LED array is matched to that of the objective [12]. For single-frame autofocusing, we turn on a red and a green LED to illuminate the sample from two different incident angles (Fig. 1(a)). If the sample is placed at an out-of-focus position, there would be a separation between the red and green channels of the sample (Fig. 3(a1) and 3(a3)). Our autofocusing scheme is to identify the separation between the red and green channels and then recover the defocus distance based on this separation. We choose red and green colors because they generate better contrast for regular hematoxylin and eosin (H&E) stained tissue slides. We use an illumination NA of ~0.4 for red and green LED illuminations. A larger illumination NA leads to a larger image difference between the red and green channels. A smaller illumination NA, on the other hand, leads to a smaller separation of the two copies.

Figure 3(a) shows the captured images under red and green illumination. Figure 3(b) shows the measured calibration curve between the two-copy separation and the lens ring position. We use a gradient descent optimization process to recover the separation between the red and green channels shown in Fig. 3(a1) and 3(a3). This optimization process calculates the gradient of the mutual information metric in the direction of the extrema in each loop [14]. We sample ~10,000 pixels in the captured images and use 5-10 loops for the optimization process. The time to converge is ~0.05 seconds. We have also tested the autofocusing performance on 1500 tiles of 5 different samples. The averaged focusing error is ~0.33 µm.

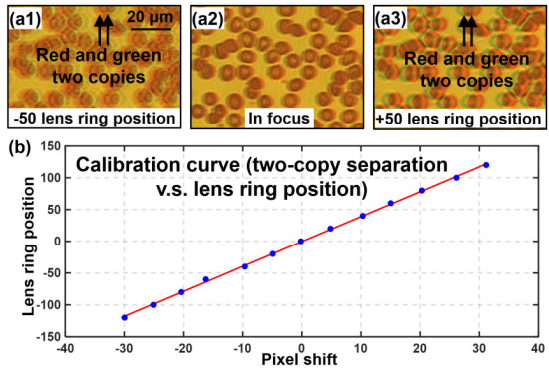

Fig. 3. Single-frame autofocusing scheme. (a) Captured images under red and green LED illumination. The separation of the red and green copies can be used to recover the defocus position of the sample. (b) The calibration curve between the two-copy separation and the lens ring position.

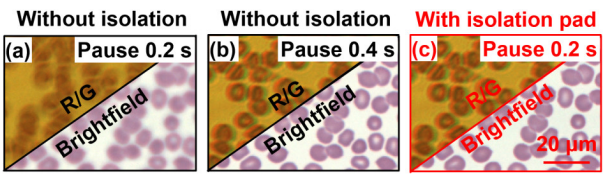

Fig. 4. Reducing system vibration via the isolation pad. The R/G and corresponding brightfield images captured at 0.2 s (a) and 0.4 s (b) after the actuation of the y stage, with no isolation pad. (c) The images captured at 0.2 s after the actuation of the y stage, with the isolation pad.

In the reported system, another critical consideration is to enable fast stage actuation. Fast stage actuation, however, would lead to vibration of the system and generate errors for defocus distance calculation. Figure 4(a) shows the captured R/G image at 0.2 s after the actuation of the y stage. The vibration leads to some random motion blur in Fig. 4(a). The resulting defocus distance calculation is, thus, not correct, and the captured brightfield image is out-of-focus. Figure 4(b) shows the images captured at 0.4 s after the stage actuation. In this case, the vibration has been settled down and the resulting defocus distance calculation is correct for the brightfield image. In our implementation, we reduce the vibration by placing a Sorbothane isolation pad under the y-scanning stage in Fig. 1(b2). With this isolation pad, the vibration has been significantly reduced. Figure 4(c) shows the captured R/G and brightfield images at 0.2 s after the stage actuation.

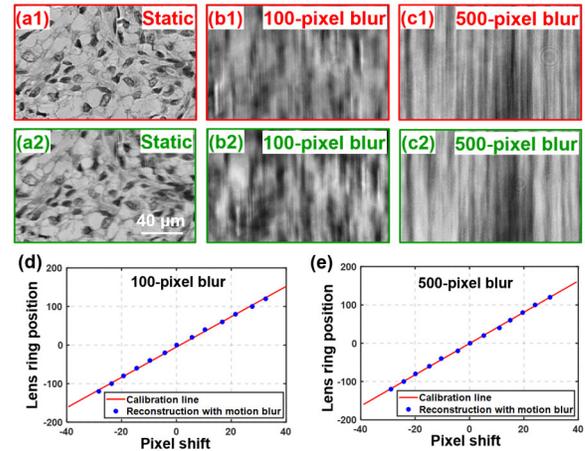

Fig. 5. Robustness to the motion blur along the y-direction. The captured red (a1) and green channels (a2) at the static state. The cases with 100-pixel motion blur (b), and 500-pixel motion blur (c). The measured two-copy separation for 100-pixel motion blur (d) and 500-pixel motion blur (e).

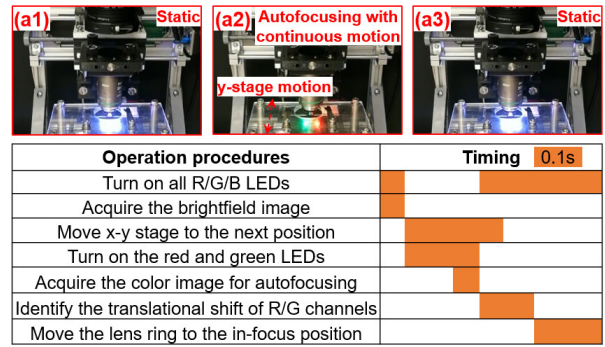

Fig. 6. (Visualization 3) Operation procedures for the OpenWSI platform. In between two static brightfield image acquisitions (a1) and (a3), we perform autofocusing with continuous y-stage motion (a2).

The autofocusing method employed in the reported system is not sensitive to the motion blur along the y-direction, which is perpendicular to the direction of two-copy separation. Figures 5(a-c) show the red and green channel images with 0-, 100-, and 500-pixel motion blurs. Figure 5(d) and 5(e) show that the recovered defocus positions are in a good agreement with the calibration curve under the y-motion blurs.

Figure 6 and Visualization 3 show the operation procedure for WSI, where we perform autofocusing with continuous y-stage motion (Fig. 6(a2)). This autofocusing process is performed in between two static brightfield image acquisitions in Fig. 6(a1) and 6(a3).

In our system, there are minor pincushion distortions at the edge of the captured image (image magnification slightly increases with the distance from the center). The pincushion distortions lead to stitching errors in Fig. 7(a). We use the following procedures to digitally correct the pincushion distortions. First, we use a hole-array mask to measure the pincushion distortion. Inset of Fig. 7(a) shows the distorted hole positions and the corresponding ground-truth positions. Second, we create a mapping equation to map the distorted hole positions to the ground-truth positions (inset of Fig. 7(b)). Third, the mapping equation is applied to the captured brightfield image. The processing time for distortion correction is ~0.13 s for each image and it can be implemented in parallel with the image acquisition process. Figure 8 shows a sample whole slide image captured using the reported platform. The lens ring position over the entire field of view is shown in Fig. 8(a) and the whole slide image is shown in Fig. 8(b). The acquisition time for this 10 mm by 11 mm sample image is 63 s (Visualization 3). For a 225-mm$^2$ area, the acquisition time is ~2 mins and the imaging throughput is comparable to existing high-end platforms.

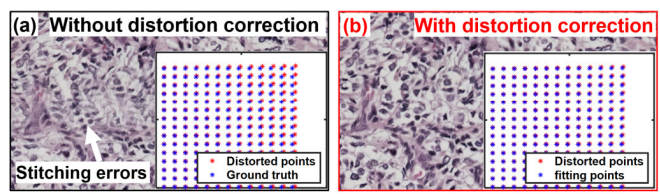

Fig. 7. Pincushion distortion correction. (a) Image stitching errors due to pincushion distortion. (b) No errors after digital distortion correction.

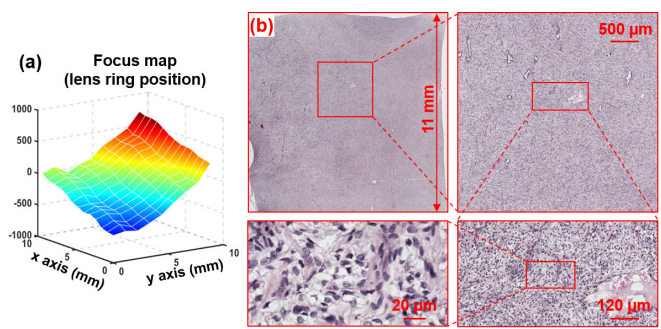

Fig. 8. Testing the OpenWSI system for whole slide imaging. (a) The generated focus map during the scanning process. (b) The captured whole slide image of a kidney cancer section. The acquisition time is 63 s for this 11 mm by 10 mm sample. We use ImageJ Fiji plugin for image stitching (refer to http://www.gigapan.com/gigapans/218291).

In summary, we report a low-cost, high-throughput WSI system termed OpenWSI. It offers a turnkey solution to transform the high-end WSI platforms into one that can be made broadly available and utilizable without loss of capacity. From the technology point of view, the reported system has several advantages compared to conventional WSI systems. 1) It employs a real-time autofocusing strategy that allows the system to be built with a low-cost CNC router. The autofocusing process can be performed with continuous sample motion. 2) It performs remote focus control at the image plane via the ultrasonic motor ring. This strategy allows us to employ a cost-effective photographic lens for 3D imaging and rapid focus tracking. 3) Multiple modalities such as phase-contrast, Fourier ptychographic, and 3D tomographic imaging can also be integrated into the reported system via the LED array. One future direction is to integrate the reported platform with an automatic slide loading system.

From the application point of view, the impacts of the reported platform are far-reaching as high-content images are desired in many fields of biomedical research as well as in clinical applications. The dissemination of the proposed platform in an affordable budget for individual researchers could lead to new types of experimental designs in small labs. In the medical realm, one strategy taken by the National Cancer Moonshot initiative to fight cancer cooperatively is to create an image database for different cases and connect scientists and pathologists for online collaboration. Such a database would allow researchers to find similarities in cancer and perform tissue driven data mining to find a cure. Converting the tissue sections and various biological samples into high-content images is the first step in this strategy. The reported OpenWSI platform holds the potential to address the challenges of high-throughput imaging and allow individual pathologists to use the WSI system.

The following 7 files for the OpenWSI system can be downloaded at [15]: 1) part list, 2) instruction on customized parts, 3) focus control of the Canon lens, 4) resolution test, 5) digital distortion correction, 6) 3D design files, and 7) demo code.

**Funding.** This work is supported by the UConn SPARK fund.

**Disclosures.** G. Z. has conflicts of interest with Pathware and Instant Imaging Technology, which did not support this work.